\documentclass[twocolumn]{aastex631}
\usepackage{amsmath}
\usepackage{CJK}
\usepackage{multirow}

\newcommand{\dechms}[4]{$#1^{\rm h}#2^{\rm m}#3\mbox{$^{\rm s}\mskip-7.6mu.\,$}#4$} 
 
\newcommand{\decdms}[4]{$#1^{\circ}#2'#3\mbox{$''\mskip-7.6mu.\,$}#4$}

\shorttitle{Eccentric dust ring in IRS 48}
\shortauthors{Yang et al.}
\graphicspath{{./}{figures/}}

\begin{document}
\begin{CJK*}{UTF8}{gbsn}

\title{Eccentric Dust Ring in the IRS 48 Transition Disk}

\correspondingauthor{Haifeng Yang}
\email{hfyang@pku.edu.cn}

\author[0000-0002-8537-6669]{Haifeng Yang (杨海峰)}
\altaffiliation{Boya Fellow}
\affil{Kavli Institute for Astronomy and Astrophysics, Peking University, Yi He Yuan Lu 5, Haidian Qu, Beijing 100871, People's Republic of China}

\author[0000-0001-5811-0454]{Manuel Fern\'andez-L\'opez}
\affiliation{Instituto Argentino de Radioastronom\'ia (CCT-La Plata, CONICET; CICPBA), C.C. No. 5, 1894, Villa Elisa, Buenos Aires, Argentina}

\author[0000-0002-7402-6487]{Zhi-Yun Li}
\affiliation{Department of Astronomy, University of Virginia, Charlottesville, VA 22903, USA}

\author{Ian W. Stephens}
\affiliation{Department of Earth, Environment, and Physics, Worcester State University, Worcester, MA 01602, USA}

\author{Leslie W. Looney}
\affiliation{Department of Astronomy, University of Illinois, 1002 West Green Street, Urbana, IL 61801, USA}

\author[0000-0001-7233-4171]{Zhe-Yu Daniel Lin}
\affiliation{Department of Astronomy, University of Virginia, Charlottesville, VA 22903, USA}

\author{Rachel Harrison}
\affiliation{Department of Astronomy, University of Illinois, 1002 West Green Street, Urbana, IL 61801, USA}

\begin{abstract}
Crescent-shaped structures in transition disks hold the key to studying the putative companions to the central stars.
The dust dynamics, especially that of different grain sizes, is important to understanding the role
of pressure bumps in planet formation. In this work, we present deep dust continuum observation with high
resolution towards the Oph IRS 48 system. 
For the first time, we are able to significantly trace and detect emission along $95\%$ of the ring crossing the
crescent-shaped structure. The ring is highly eccentric with an eccentricity
of $0.27$. The flux density contrast between the peak of the flux and its counter part along the ring is 
$\sim 270$.
In addition, we detect a compact emission toward the central star. If the emission is an inner circumstellar disk inside the cavity, it has a radius of at most a couple of astronomical units with a dust mass of 
$1.5\times 10^{-8}\rm\, M_\odot$, or $0.005\rm\, M_\oplus$. 
We also discuss the implications of the potential eccentric orbit on the proper motion of the crescent, the putative secondary companion, and the asymmetry in velocity maps.
\end{abstract}

\section{Introduction}

Transition disks (TD) are protoplanetary disks with large inner cavities.
These cleared inner regions hint at the existence of a companion (e.g. \citealt{Marsh1992}) 
or a history of photoevaporation (e.g. \citealt{Alexander2014} and references therein). 
The Atacama Large Millimeter/submillimeter Array (ALMA) has revealed many transition disks with 
diverse structures (see \citealt{vanderMarel2021_asymmetry} and reference therein). 
Among these structures, the crescent-shaped structures are of
particular interest, because they directly link to the 
putative companion. 
Theoretically, the creation of such structures have at least three possible origins.
A crescent can be a long-lived vortex caused by 
Rossby wave instability (RWI) \citep{Zhu2014} or a dust horseshoe from the 
overdensity at the cavity edge \citep{Ragusa2017}. 
The crescent-shaped structures from these two 
mechanisms are triggered by companions but of different
masses \citep{Dong2018,Ragusa2020}. They move at the local Keplerian speed and both 
cause azimuthal dust segregation \citep{Birnstiel2013}. 
The third mechanism relies on the eccentric disk caused by a massive companion \citep{Ataiee2013,Kley2006}. 
In this case, an eccentric disk induced by a companion has an overdense region near the apocenter,
which manifests itself as a slowly precessing crescent-shaped structure with a negligible proper 
motion.

Among all transition disks, Oph IRS 48 stands out and draws interest and studies for several reasons.
It has a prominent crescent-shaped structure with a density contrast of $>100$, but only at 
(sub)millimeter wavelengths \citep{vanderMarel2013}. 
On the contrary, the mid-IR and $^{12}$CO line emissions both show symmetric structures \citep{vanderMarel2013}.
In addition, the azimuthal concentration increases towards longer wavelengths \citep{vanderMarel2015}, hinting at dust segregation of different grain sizes,
which supports the vortex picture \citep{Zhu2014}. 

In addition to the crescent-shaped structures, some transition 
disks may have hot dust near the central star or even resolved
inner disks. They have a substantial infrared excess and are classified as pre-transitional
disks (PTD), an intermediate state between full disks and transitional disks \citep{Espaillat2010,Espaillat2014}
\footnote{Recently, \cite{Francis2020} showed that there is no clear correlation between NIR excess and central mm-dust emission. The PTD/TD classification is currently under debate.}. 
Whether IRS 48 is PTD or TD is uncertain due to the presence of strong PAH emission \citep{Geers2007}, even though it has $3.8\%$ NIR excess \citep{Francis2020}. 
Previous observations didn't resolve the putative inner disk associated with the infrared
excess, and gave an upper limit of the dust mass of $0.009\rm\,M_\oplus$ \citep{Francis2020}.
At the same time, IRS 48 has an appreciable mass accretion rate of $10^{-8.4}\rm\, 
M_\odot/yr$ \citep{Salyk2013}.
A detection of the inner disk in the IRS 48 system will confirm its classification 
as a PTD, and help us understand the evolution of transition disks.

In this work, we present new deep observations towards IRS 48 with high resolution. 
The structure of the paper is as follows. In Sec.~\ref{sec:obs}, we discuss the observation and data 
reduction. In Sec.~\ref{sec:res}, we discuss the main features of the data: an eccentric ring and
the detection of dust emission inside the cavity. 
In Sec.~\ref{sec:discussion}, we discuss the proper motion and the physics behind the eccentric ring. 
We present our conclusions in Sec.~\ref{sec:summary}. 

\section{Observations}
\label{sec:obs}

Observations were conducted on 2021 June 7, June 14 and July 19 using ALMA Band 7 (0.87~mm) under the project code
2019.1.01059.S (PI: H. Yang). ALMA used 42-46 antennas in six execution blocks (approximately 1.75 hours each) in two different array configurations (C43-6 and C43-7), which together provided baselines ranging from 15 to 3700~m. Weather conditions were good for
0.87\,mm observations. The mean precipitable water vapor column ranged between 0.6 and 0.9\,mm, and the system
temperature were between 132 and 171 K. The
experiment was primarily designed for studying the polarization of the dust emission toward IRS\,48. Hence, we
tuned four ALMA basebands dedicated for the dust continuum emission, centered at 336.5, 338.4, 348.5 and 350.5\,GHz,
all having a nominal 2.0\,GHz bandwidth. The observations toward IRS\,48 were intertwined with visits to the phase and 
the polarization calibrators every $\sim4$ and $\sim40$ minutes, respectively. The observations include also periodic 
visits to a check source (quasar J1647-2912) every 15 minutes. The total integrated time over IRS\,48 was 3.8 hours, 
and there was sufficient parallactic angle coverage for polarization calibration. The phase center was located at ($\alpha$,$\delta$)$_{ICRS}
$=(\dechms{16}{27}{37}{190},\decdms{-24}{30}{35}{030}). The calibration by the ALMA staff was produced using the 
Common Astronomy Software Applications (CASA) package version v6.2.1.7 \citep{CASA} in the delivered data. 
J1337-1257 and J1517-2422 were used as the flux and band-pass calibrators on different days; J1700-2610 and J1647-2912 
were the phase calibrators (average fluxes of 0.94 and 0.097\,Jy, respectively); J1733-1304 was the polarization 
calibrator in all execution blocks. ALMA Band\,7 observations have a typical absolute flux uncertainty of 10\%, and 
the polarization uncertainties are usually constraint by the gain leakages, which are less than 5\%. The images of the continuum were also made using CASA. To 
construct them we combined the four continuum basebands avoiding some spectral channels with potential line emission (C$^{17}$O 
(3-2) and several CH$_3$OH transitions). We ran two phase-only self-calibration iterations on the continuum Stokes\,I 
data. The solution intervals used for the first and second were infinite  (i.e., a solution interval over the whole dataset) and 25 seconds, respectively. The final 
signal-to-noise ratio of the Stokes\,I image is just above 1250. The selfcalibration solutions from the continuum Stokes\,I 
were then applied to the continuum Stokes\,QUV. To clean the images, we use the CASA task \texttt{tclean} using the Hogbom 
algorithm with a Briggs weighting of 0.5. The final continuum synthesized beam is $0\farcs{11}\times0\farcs{072}$, with a position angle of
$-73\degr$. 
Note that, in this case, the self-calibration did not affect the positional accuracy. The difference between the measured peak positions before and after self-calibration are well within astrometric accuracy.
The rms noise level measured in the Stokes\,I image is 14\,$\mu$Jy. For the Stokes\,QUV images the rms 
noise level is 12\,$\mu$Jy. 
The Letter focuses on the features in the Stokes I data; the polarization data will be discussed elsewhere.  

We use the parallax, the proper motion, and the location of the IRS 48 star from Gaia DR3 \citep{GaiaDR3}. 
From the parallax, we derive a distance of $136$ pc, which is slightly different from
the distance, $134$ pc, inferred from Gaia DR2 data \citep{vanderMarel2021_IRS48,GaiaDR2}. 
According the Gaia DR3, in the year 2016, the star is at 
(\dechms{16}{27}{37}{180},\decdms{-24}{30}{35}{416}).
It has an proper motion of $-8.72\rm\, mas/yr$ in RA and $-24.4\rm\, mas/yr$ in Dec. 
We derived the location at the time of our observation as (\dechms{16}{27}{37}{177},\decdms{-24}{30}{35}{550}),
assuming $5.5$ years time difference. This will be the center of all images in this paper.

\section{Results}
\label{sec:res}

\begin{figure*}[!hbt]
\includegraphics[width=0.48\textwidth]{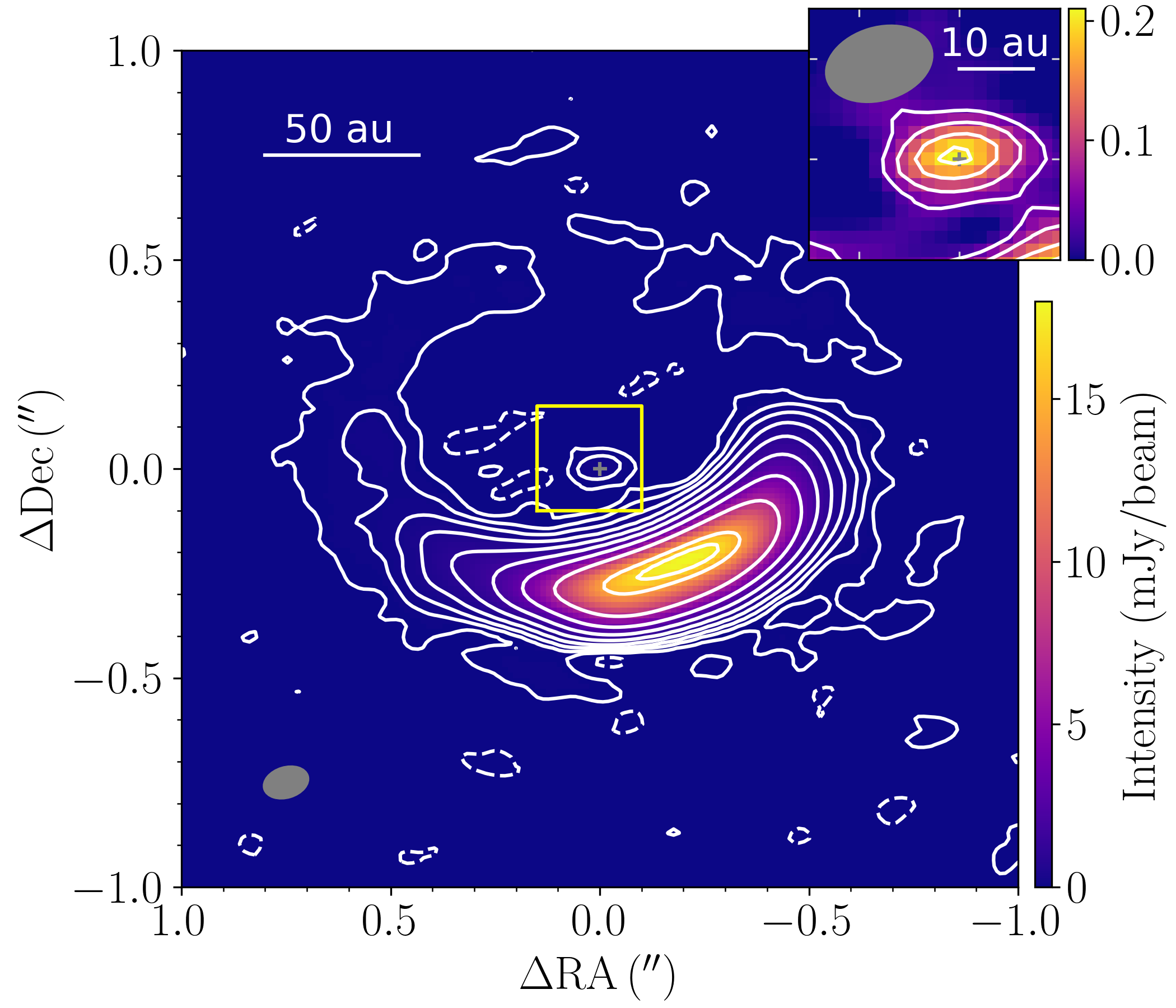}
\includegraphics[width=0.48\textwidth]{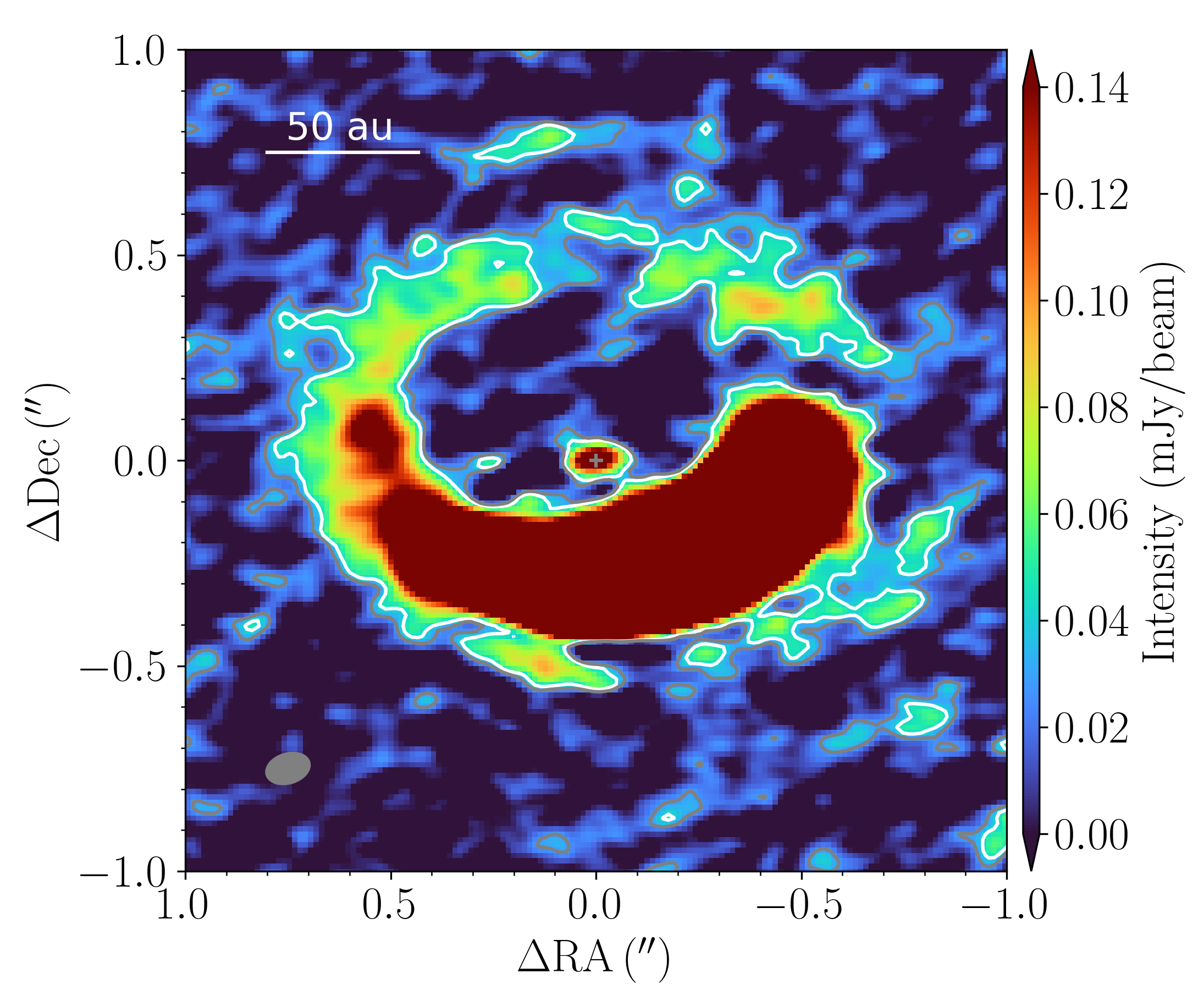}
\caption{
\textit{Left:} The synthesized dust continuum image of IRS 48 at $870\rm\, \mu m$ or $345$ GHz. Both the 
colormap and the contours represent the flux density in mJy/beam. The contours are 
plotted at the levels of $(-3, 3, 8, 16, 64, 128, 256, 512, 1024, 1200)\times \sigma$, where 
$\sigma=14\rm\, \mu Jy/beam$ is the rms noise. The synthesized beam is $0\farcs11 \times 0\farcs072$.
The inset shows a $0\farcs15$ wide zoom, with contours representing the flux density at the 
levels of $(3, 6, 10, 14)\times \sigma$.
The star location is labeled as a gray cross.
\textit{Right:} The same image but with a different colormap with a maximum value
$0.14\rm\, mJy/beam$, $10\sigma$, to saturate the crescent-shaped structure and 
display the dim long tail better. 
The white and gray contours correspond to flux densities of 
$3\sigma$ and $2\sigma$, respectively.}
\label{fig:image}
\end{figure*}

The primary-beam-corrected dust continuum image from our observation is shown in 
Fig.~\ref{fig:image}. The peak of the emission has $1285$ S/N or $18.00\rm\, mJy/beam$. In addition 
to the well-known crescent-shaped structure, we also detect a 
long tail of dust emission trailing behind the
crescent-shaped structure, with respect to the 
counter-clockwise rotation of the disk \citep{Bruderer2014}, and some diffuse 
emission with over $3\sigma$ detection in the north west part of the disk. These two 
structures form an ellipse around the central object and will be discussed in Sec.~\ref{ssec:ellipse}.
We also detect some emission at the $15\sigma$ level near the central object and separate it from the
outer crescent for the first time. We will discuss this emission in Sec.~\ref{ssec:innerdisk}.

\subsection{Eccentric ring}
\label{ssec:ellipse}
\begin{figure*} [!hbt]
\centering
\includegraphics[width=\textwidth]{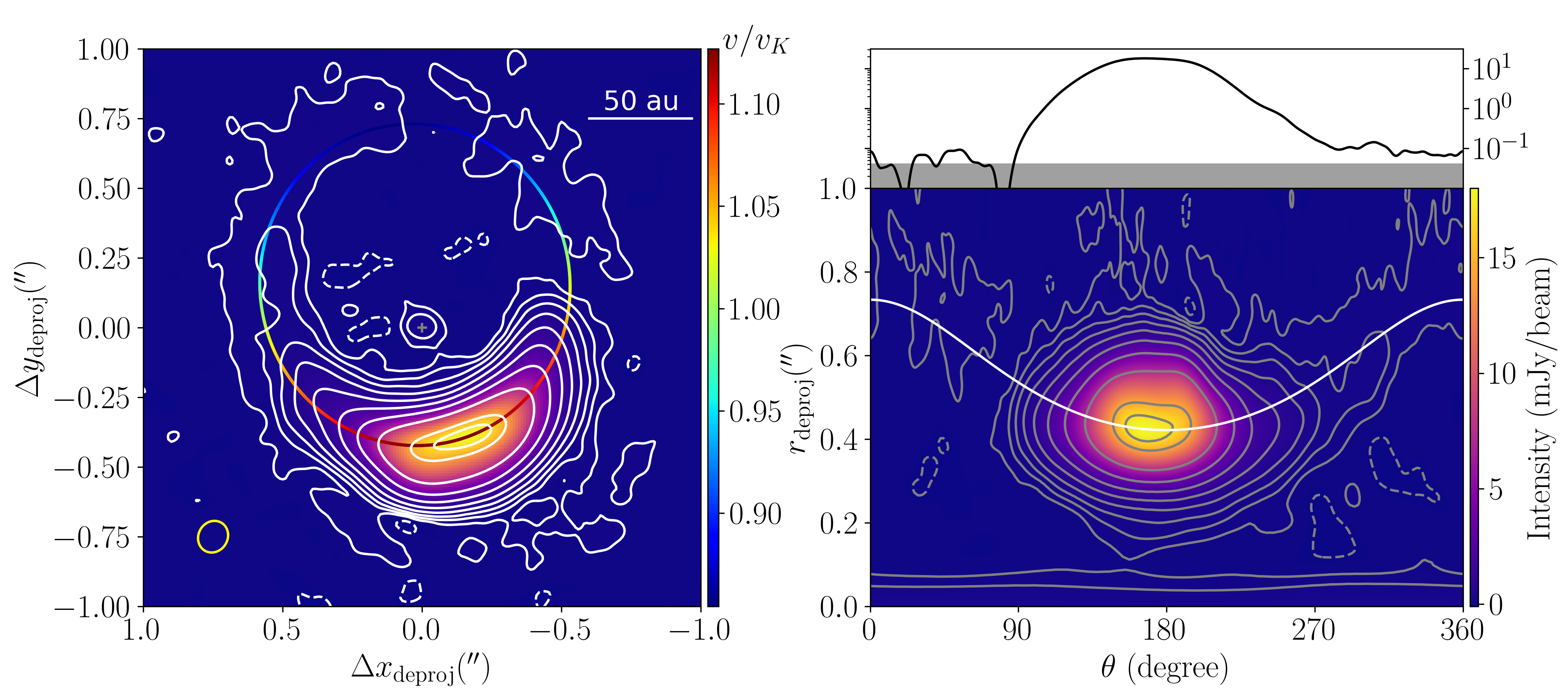}
\caption{\textit{Left:} The fitted elliptical ring in the deprojected view. 
The image shares the colorbar with the bottom right panel.
The central star (gray cross; derived from Gaia DR3) is 
assumed to be one of the foci. The contours are plotted at the same levels as in Fig.~\ref{fig:image}. 
The colored ellipse shows the best fitted ring, with colors representing the ratio of the orbital velocity to
the local Keplerian velocity, $v/v_K$ on the colorbar. 
The yellow ellipse at the lower left corner shows the synthetic beam deprojected to the disk plane.
\textit{Bottom Right:} The deprojected image in $(r, \theta)$ coordinates, where $\theta=0$ corresponds to the aphelion of the best fitted elliptical ring.
Note that a circular orbit is a straight horizontal line in this
view.
\textit{Top Right:} The flux density along the best fitted elliptical ring in logarithmatic scale. 
It shares the x-axis with the panel below. 
The gray contours are plotted at the same levels as in Fig.~\ref{fig:image}.}
\label{fig:ellipse_fit}
\end{figure*}

Our ALMA observations are the deepest high resolution (sub)mm observations towards IRS 48 to date, reaching a noise level of $14\rm\, \mu Jy/beam$ at an angular resolution of $0\farcs11\times 0\farcs072$. For the
first time we are able to significantly trace and detect emission from about $95\%$ of the ring crossing the crescent-shaped
structure. Assuming an inclination angle of $50^\circ$ and a position angle of $100^\circ$ 
\citep{Bruderer2014}, we deproject the image to the disk plane. The results are shown in 
the left panel of Fig.~\ref{fig:ellipse_fit}. We can see that the north-northwest tail behind the crescent-shaped
structure and the diffuse emission in the 
northwest part of the disk structure
form an elliptical pattern rather than a circular pattern. 

To fit the elliptical ring, we first parameterize the ellipse with the semi-major axis $a$, 
the eccentricity $e$, and the position angle of the major axis of the ellipse. We fix one of the foci
on the central star. The fitting is done on the deprojected image, and the loss function is defined as 
$\mathcal{L}\equiv \sum I(x,y) f(dr(x,y))$, where $I(x,y)$ is the intensity and $dr(x,y)$ is the difference
of the radial distance from the center between the point in the image and the target ellipse. 
The $f(dr)$ function is defined as:
\begin{equation}
f(dr)=\left\{
\begin{array}{ll}
(dr/\delta_p)^2   & \qquad dr\le 0 \\
(1.5dr/\delta_p)^2 & \qquad dr>0
\end{array}
\right.,
\end{equation}
where $\delta_p=0\farcs02$ is about the pixel size of our image, or $1/5$ of the beam size. The choice of 
$\delta_p$ and the pre-factor are arbitrary as we minimize the loss function to get the best fitted model.
The loss function is engineered to have our fitted ellipse crossing the central peak of the crescent-shaped structure, 
as the one we present in the next paragraph.

The best fit is plotted as a colored ellipse in Fig.~\ref{fig:ellipse_fit} with the color representing
the ratio of the orbital velocity to the local Keplerian velocity\footnote{Given semi-major axis $a$ and the distance to the center $r$, the ratio of the orbital velocity along elliptical orbit to the Keplerian velocity assuming circular orbit with a radius of $r$ is $v/v_K=\sqrt{2-(r/a)^2}$.} $v/v_K$. The best fit has a semi-major axis $a=0.57''$,
or $78$ au, and an eccentricity $e=0.27$. At the perihelion, the velocity is about $1.13v_K$, which is slightly 
super Keplerian. 
We will discuss the implication of the elliptical dust distribution in more detail in Sec.~\ref{ssec:pm}.

In the bottom right panel, we plot the same deprojected image in the $(r, \theta)$ coordinates. The $\theta$
is defined such that the perihelion corresponds to $\theta=180^\circ$. Note that a circular
orbit is a horizontal straight line in these coordinates. Our fitted ellipse is very eccentric with 
an aphelion to perihelion distance ratio of $1.78$. 
In the top right panel, we plot the flux density 
profile in logarithmic scale along the fitted ellipse. We can see that the density structure is
not symmetric with respect to the peak, and it resembles a droplet or a tadpole. The head is rounder with
about $90^\circ$ degree spread in azimuthal angle whereas the tail spreads over almost $180^\circ$. The diffuse emission in the
northwest part of the disk has an azimuthal extent of $40$-$50^\circ$.
Since the emission along part of the ellipse is not robustly detected, the ratio of the maximum intensity is at least 400 times the minimum. 
As a reference for interested readers, the flux density contrast of the
peak with the point along the ring with $180^\circ$ position angle difference is $271$.

The necessity for the ring to be eccentric is clear given that the north part is $\sim 1.8$ times further from 
the central star than the peak, and the exact value of the inclination angles does not change the
eccentricity too much, as long as the position angle of the 
disk is fixed. 
From the left panel of Fig.~\ref{fig:ellipse_fit}, we 
can see that the major axis of the ellipse is very close
to being vertical. In this case, changing the inclination angle
will change both the apocenter and pericenter distances by 
the same factor, leaving the ratio unchanged as $\sim 1.8$. 
The semi-major axis will be different for the new inclination
angle, but the eccentricity can be inferred directly as
$e\approx (1.8-1)/(1.8+1)\approx 0.28$. 
So the eccentricity does not depend on the adopted 
inclination angle too much. 
The assumption of the central star as one focus has a much
larger impact on the eccentricity. We will discuss
these possibilities in more detail in Sec.~\ref{ssec:manyfocus}.

\subsection{Detection of central emission}
\label{ssec:innerdisk}
Figure \ref{fig:image} clearly shows a central source with at least $3\sigma$ emission over 
more than $2$ beams. This source has a peak flux density of $0.21\rm\, mJy/beam$, at a signal-to-noise level $15\sigma$. We conduct a simple 2D-Gaussian fit to this central continuum emission. The center is only $3\rm\,mas$  west of the
central star, well within one pixel of our image ($20$ mas). The spatial coincidence between the star 
location inferred from the Gaia data and the center of the central emission adds confidence to the 
location determination to both of them. It also makes it unlikely for this source to be either background contamination 
or random calibration or cleaning artifact. The major and minor axis of the 2D-Gaussian fit is $0.095''\times 0.052''$, which is smaller 
than our beam size. Even though the $3\sigma$ contour is larger than 2 beams, our data is still in 
agreement with a point source.

The simplest assumption is that this central emission comes from dust. However, before calculating a dust mass about the central source, we explore whether this emission may come from other sources.
Firstly, the central emission cannot be completely explained by emission coming from a central star. 
The star IRS 48 was estimated as a $2\rm\, M_\odot$ star using kinematic modeling \cite{Brown2012}. 
The effective temperature and the bolometric luminosity was estimated as $9520$ K and $17.8\rm\, L_\odot$ (\citealt{Brown2012}; rescaled by \cite{Francis2020} using new distance from Gaia DR2 data \citep{GaiaDR2}).
The effective temperature and the bolometric luminosity combined yield a 
radius of $1.55\rm\, R_\odot$. 
We can then calculate the flux density at the wavelength of $870\rm\, \mu m$ as 
$7.22\times 10^{-6}\rm\, Jy$. This is $21$ times smaller than what we detect.

Another possibility to explain the central unresolved emission is that it comes 
from the free-free emission of an ionized wind. Taking the flux of the central 
source at 34 GHz ($36\rm\, \mu$Jy, \citealt{vanderMarel2015}) and the flux derived in 
our 2D-Gaussian fit at 343.5 GHz ($221\rm\, \mu$Jy), we obtain a spectral index of 
$0.78\pm0.20$. This spectral index is consistent with the partially optically thick 
free-free emission of a jet/wind. Although ionized jets are common among young 
stars the turnover frequency from the partially optically thick to the optically 
thin regimes depends upon the ionized gas density at the jet base and it is usually 
located at the centimeter part of the continuum spectrum \citep[e.g.,][]{1986Reynolds,2022Mohan}. 
Therefore, if the emission of the central source at 343.5 GHz has a 
contribution from a thermal ionized jet, it would most probably be in the optically 
thin regime, with a spectral index of $-0.1$. Assuming that the emission of a 
thermal jet is optically thin from 34 to 343.5 GHz, the free-free contribution 
would be $29 \mu$Jy at 343.5 GHz, roughly 10 times smaller than the measured flux.

Neither the emission from the protostellar photosphere, nor the emission from an 
ionized wind could alone (or added) explain the measured flux from the central 
source in the ALMA image. Still other possible contributions can be due to the 
non-thermal synchrotron emission from the protostellar magnetosphere \citep{Andre1997},
the ionized emission from the dust sublimation wall \citep[e.g.,][]{2020AnezLopez}, or 
the collisions from pebbles and planetesimals that heat the dust of a disk \citep[such as
proposed in Vega,][]{2020Matra}. The former two hypotheses are expected to be more 
prominent in the centimeter wavelength range, while the third hypothesis is somehow a more 
exotic solution that may need more theoretical work to be supported. In this work we will 
assume a more simple hypothesis, in which most of the submillimeter emission comes from the 
thermal dust emission of an inner disk. Following this idea, we can put a constraint on the 
dust mass of the central object. 
The mean flux density within the $3\sigma$ contour is about $8.65\times 10^{-5}\rm\, Jy/beam$, 
corresponding to a brightness temperature of $0.164$ K. This is extremely optically thin for
 any reasonable dust temperature. 
The area of this region is measured as $291\rm\, au^2$. Given the small size of the central source (maximum distance from the central star is $\sim 10$ au), we assume a dust temperature of $100$ K. We also assume a dust opacity of $3.5\rm\, cm^2/g$, which is the fiducial dust model from \cite{Birnstiel2018} with $1$ mm maximum grain size at $870\rm\, \mu m$ wavelength. 
Under these assumptions, the dust mass is estimated to be about $1.5\times 10^{-8} M_\odot$, 
or $0.005\rm\, M_\oplus$. 

\section{Discussion}
\label{sec:discussion}
\subsection{Proper motion}
\label{ssec:pm}

\begin{figure*}
\centering
    \includegraphics[width=0.45\textwidth]{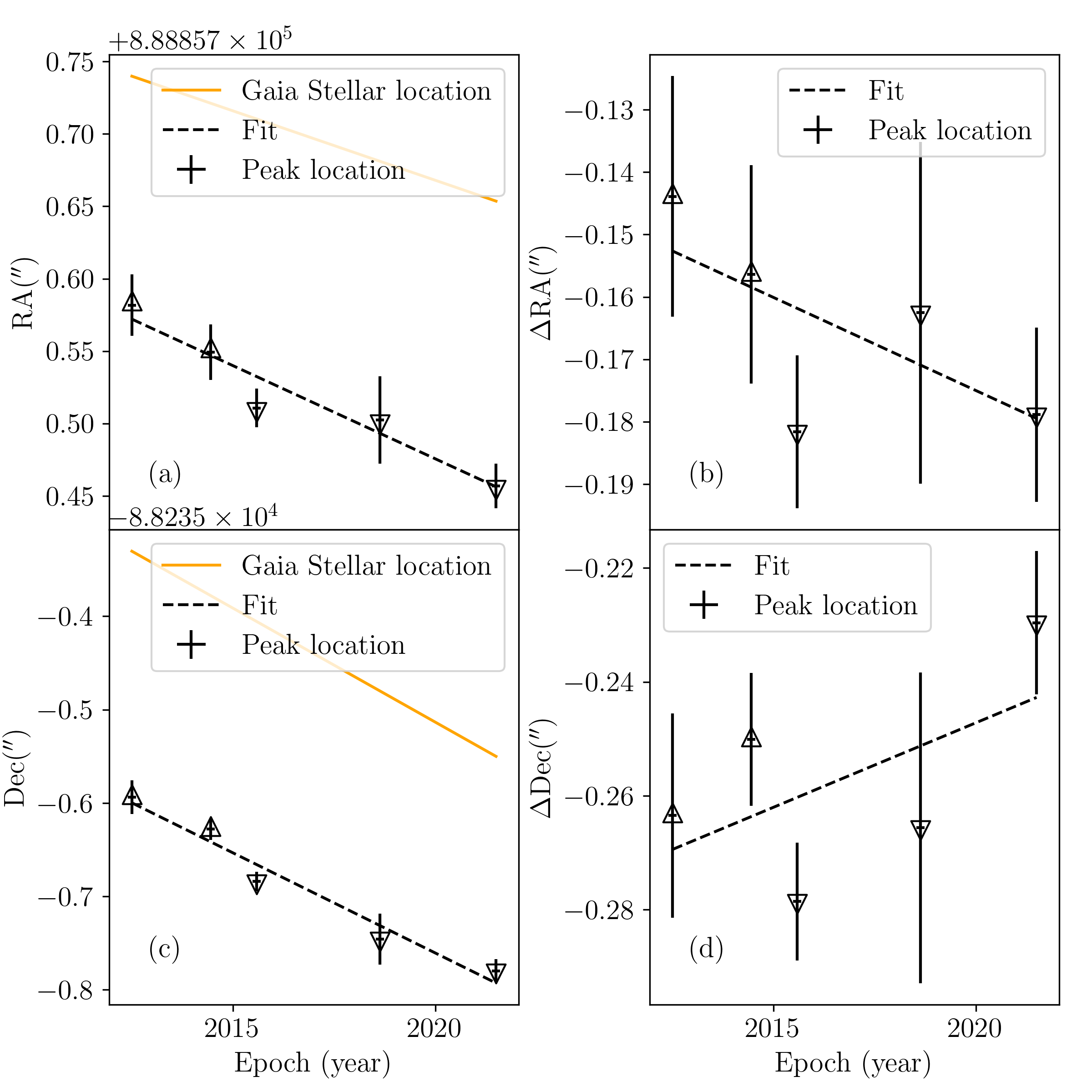}
    \includegraphics[width=0.48\textwidth]{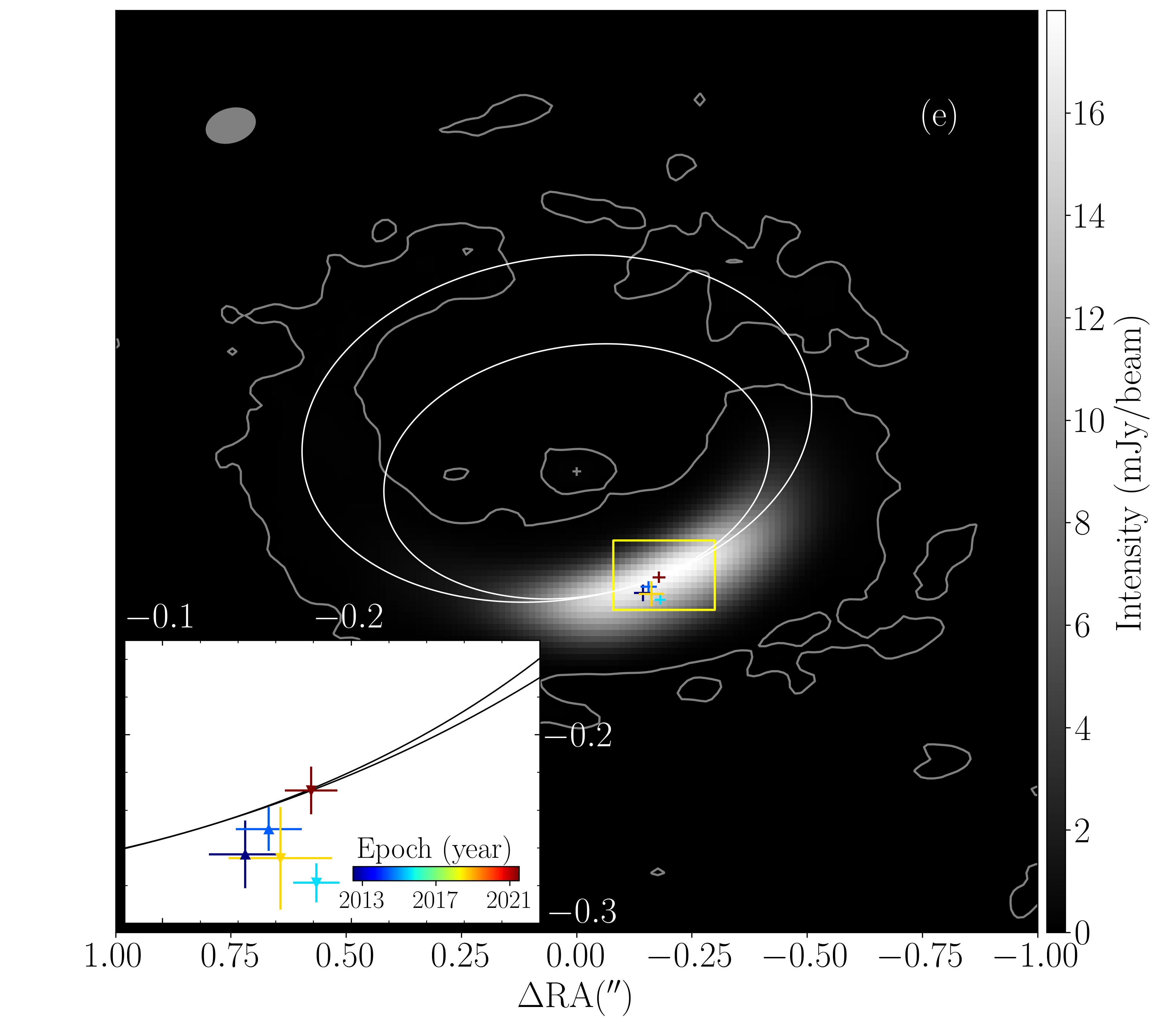}
    \caption{The Right Ascension (a) and Declination (c) of the crescent peak from archival ALMA data and
    the star assuming Gaia proper motion in units of arcsec. 
    The relative displacement from the assumed stellar location at 
    each epoch is plotted in unit of arcsec in (b) and (d). The dashed lines are linear fits to each data
    sets. The up (down) triangles represent data points observed at ALMA Band 9 (Band 7). 
    \textit{Panel (e):} The intensity image overlaid with the predicted proper motion. The smaller and 
    larger ellipses represent the circular and the best fit eccentric orbit, respectively. 
    The colorful crosses represent the fitted peak location at each epochs with errors. 
   }
    \label{fig:peak}
\end{figure*}

In Sec.~\ref{ssec:ellipse}, we fit the ring crossing the crescent-shaped structure with an
elliptical orbit having an eccentricity of $0.27$, and the current peak of the crescent-shaped structure
is near the perihelion of the orbit. Such an eccentric orbit would have a local velocity near the 
perihelion of about $1.13v_K$. The crescent-shaped structure following this orbit should also move at a super
Keplerian speed. 

Since \cite{vanderMarel2013}, the IRS 48 has been observed with ALMA at high resolution multiple times 
(\citealt{vanderMarel2015,Francis2020,vanderMarel2021_IRS48}; 2021 observations in this study).
The 9 year separation in observing time allows us to constrain the proper motion of the crescent-shaped 
structure. We obtained the archival ALMA data and use \texttt{CASA} \texttt{imfit} to conduct 2D Gaussian
fit to the crescent. 
The observation time, adopted epoch, observing band, beam size, astrometric error $\delta_\mathrm{astro}$, \texttt{imfit} results, and references are presented in Table~\ref{tbl:imfit}.
When estimating the astrometric error, we followed the ALMA Technical Handbook\footnote{
Section 10.5.2 of ALMA Technical Handbook (Cycle 9). \\ \url{https://almascience.nrao.edu/proposing/technical-handbook}
}, which is the beam size divided by the signal to noise ratio (saturates in $20$) divided by 0.9. 
For observations with resolution finer than $0\farcs15$, the positional error can be up to a factor of two higher.
Only our observations are this high resolution, so we doubled the aforementioned astrometric error to be conservative.
For the $\texttt{imfit}$ results, we present the fitted peak location translated to the ICRS frame, the error in RA and the error in Dec. 
The locations of the peaks are plotted in panels (a) and (c) in Figure~\ref{fig:peak}.
The errors in fitted RA and Dec are the square root of the sum of the squared errors from both astrometric 
error and fitting error.
We also put $0.1$ error for all epoches, which is on the order of one month. 
The stellar location from the Gaia proper motion measurement is also plotted, which allows us to calculate
the relative displacement from the central star in panels (b) and (d). We do not consider the astrometric error from Gaia.

\begin{table*}
\begin{center}
\hspace{-1em}
\begin{tabular}{lcccccccc}
\hline
Observation time & Epoch   & Band & Beam size & $\delta_\mathrm{astro}('')$ & Peak Location (ICRS) & $\delta_\mathrm{RA}('')$ & $\delta_\mathrm{Dec}('')$ & Refs \\
\hline
06/2012; 07/2012 & 2012.50 & 9 & $0\farcs32\times 0\farcs21$ & 0.018 & \dechms{16}{27}{37}{172}, \decdms{-24}{30}{35}{593} & 0.0075 & 0.0024 & 1 \\
06/2014          & 2014.45 & 9 & $0\farcs19\times 0\farcs14$ & 0.011 & \dechms{16}{27}{37}{169}, \decdms{-24}{30}{35}{628} & 0.0139 & 0.0049 & 2 \\
07/2015; 08/2015 & 2015.58 & 7 & $0\farcs18\times 0\farcs12$ & 0.010 & \dechms{16}{27}{37}{167}, \decdms{-24}{30}{35}{684} & 0.0070 & 0.0026 & 3 \\
08/2018          & 2018.63 & 7 & $0\farcs49\times 0\farcs39$ & 0.027 & \dechms{16}{27}{37}{166}, \decdms{-24}{30}{35}{745} & 0.0031 & 0.0015 & 4 \\
06/2021; 07/2021 & 2021.50 & 7 & $0\farcs11\times 0\farcs072$& 0.012 & \dechms{16}{27}{37}{163}, \decdms{-24}{30}{35}{779} & 0.0068 & 0.0030 & 5 \\
\hline
\end{tabular}
\end{center}
\caption{The proper motion of the peak of the crescent-shaped structure. $\delta_\mathrm{astro}$ is the astrometric error. $\delta_\mathrm{RA}$ and $\delta_\mathrm{Dec}$ are fitting error using \texttt{imfit} for RA and Dec, respectively. The references and ALMA project IDs are: 
(1) \cite{vanderMarel2013},2011.0.00635.SSB; 
(2) \cite{vanderMarel2015}, 2013.1.00100.S;
(3) \cite{Francis2020}, \cite{vanderMarel2021_asymmetry}, 2013.1.00100.S;
(4) \cite{Ohashi2020}, \cite{vanderMarel2021_IRS48}, 2017.1.00834.S;
(5) This work, 2019.1.01059.S.
}
\label{tbl:imfit}
\end{table*}

We note that the first two points, from \cite{vanderMarel2013} and
\cite{vanderMarel2015}, are observed at Band 9 ($440\rm\, \mu m$), whereas the others are observed at Band 7 ($870\rm\, \mu m$). 
Since the peak at $440\rm\, \mu m$ was reported to coincide with that at 9 mm \cite{vanderMarel2015}, it is 
reasonable to ignore the potential azimuthal displacement here, even though it was predicted in simulations
with self gravity \citep{Baruteau2016}\footnote{We note that the 9 mm data had a much larger beam size of $0\farcs46\times0\farcs26$. The peaks of the emission at $440\rm\, \mu m$ and at $9\rm\, mm$ could still have small non-resolved displacement.}. We will use all 5 data points to fit the proper motion for larger
sample size and time span. To highlight the difference in observing bands, we have marked the data observed at 
Band 9 and Band 7 with up triangles and down triangles, respectively, in all panels of Figure~\ref{fig:peak}.

With these caveats in mind, we fit the proper motion of the peak after subtracting the stellar proper motion as 
$(-3.0\pm1.9, 3.0\pm 2.5)\rm\, mas/yr$. 
This is 
$(-1.9\pm 1.2, 1.9\pm 1.6)\rm\, km/s$ 
at a distance of $136\rm\, pc$ and 
$(-0.35\pm 0.22, 0.35\pm 0.29)\times v_K$, 
where as $v_K=5.438\rm\, km/s$ is the Keplerian
velocity at the distance of $60$ au, the location of the peak. The magnitude of the proper motion is 
$0.49\pm 0.26\, v_K$.

In panel \textit{(e)} of Fig.~\ref{fig:peak}, we over plot the circular orbit and the best-fit 
elliptical orbit on the image. 
We also plot the peak locations as colorful dots with colors representing their observing epoch. 
We can see that 
the data points do not follow either a circular orbit or an eccentric orbit. 
There are a few possibilities that may cause this mismatch. For example, 
in the case of an undetected stellar companion with a non-negligible mass (as suggested by \citealt{Calcino2019}), the Gaia proper motions do not take into account the possible orbital motion of the primary star.
It is also possible
that the crescent peak has some epicycle motion in addition to bulk orbital motion. 
Further work with all datasets modeled more accurately and consistently 
and observations towards the IRS 48 with high resolution again in $5-10$ years may help
to understand the nature of the proper motion of the crescent peak.

\subsection{Secondary stellar companion?}
\label{ssec:manyfocus}
Theoretically, one way to drive the eccentric orbital motion is to have a secondary stellar companion.
In the simulation presented in \cite{Calcino2019}, they introduced a companion as massive as $0.4\rm\, 
M_\odot$ at the separation of $10$ au, to explain the observed asymmetry in velocity channel maps and line observations \citep{vandermarel2016}. Such a massive companion will change the mass center of the 
system significantly ($1.67$ au for their set-up), and the focus of the eccentric ring should be
displaced from the central star. 

\begin{figure}
\includegraphics[width=0.48\textwidth]{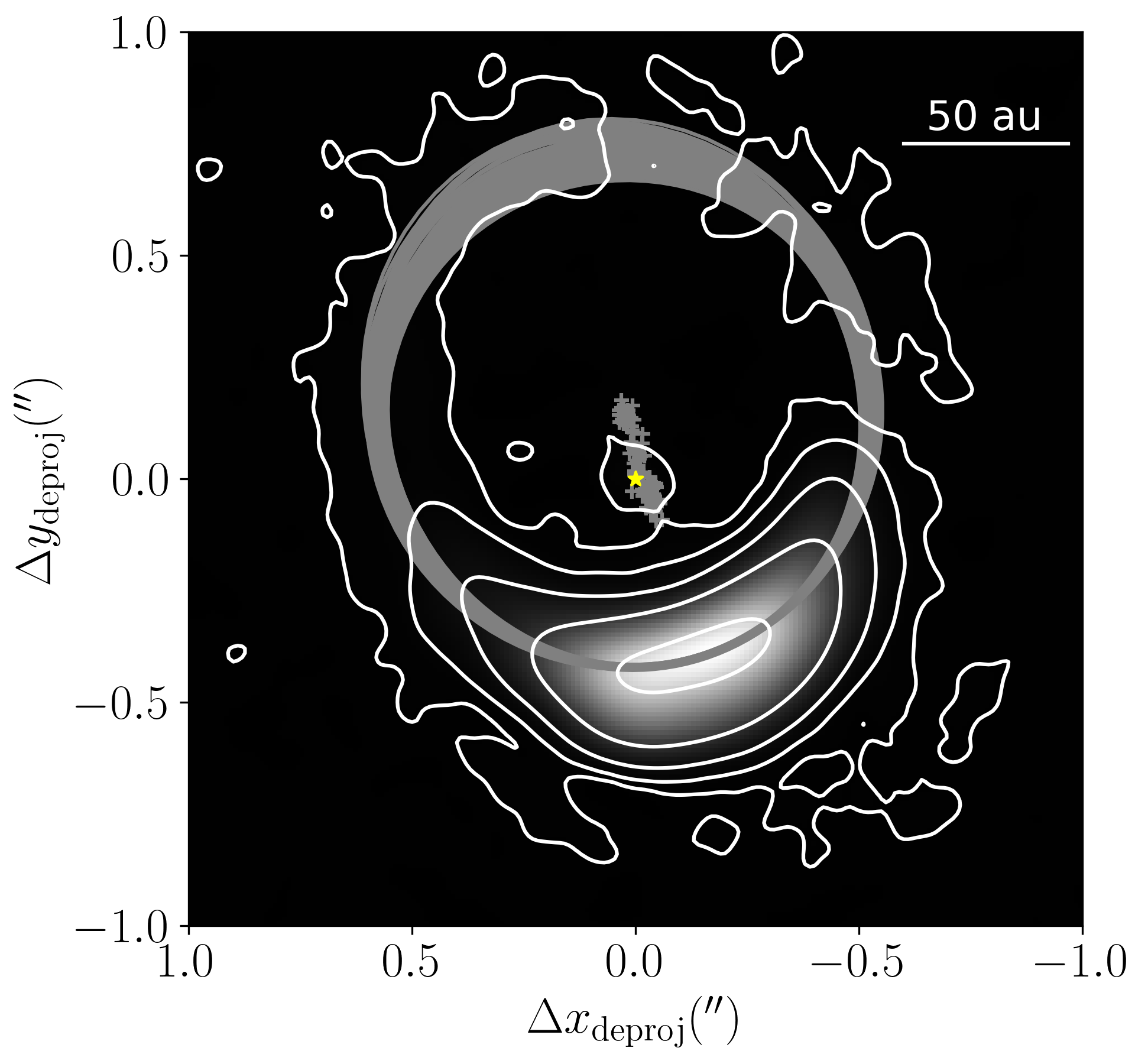}
\caption{Orbital fits shown in the deprojected plane without using the assumption of the star at one of the foci. 
We plot $1000$ gray ellipses with a reasonable fit. The worst of these orbits has a loss function that is 
only $1\%$ larger than the best. The foci (near the star) of these ellipses are also plotted as 
gray crosses. The yellow star at the center marks the primary star location.}
\label{fig:new_fit}
\end{figure}

To explore these possibilities, we relax the constraint in Sec.~\ref{ssec:ellipse} and introduce the 
location of the focus as two additional parameters. The loss function is still defined with the 
difference between the distances towards the new focus as its argument. There are many ellipses with 
similar levels of loss functions. In Fig.~\ref{fig:new_fit}, we plot $1000$ orbits and their foci
with reasonable fits to our data. 
Among these ellipses, the largest loss function is only $\sim1\%$ larger than the 
smallest one. Despite the similarity in their loss functions, the location of the foci differs by
almost $50$ au. 

The uncertainty of the fitted focus mostly comes from the dispersed nature of the emission. Deeper 
observations with lower noise are unlikely to give much better constraints. 
Aside from constraints from the dust emission, one can search for the potential 
secondary star with astrometry, in addition to the existing constraint from photometry (\citealt{Wright2015}, c.f. discussions by \citealt{Calcino2019}). 
If there is a massive secondary stellar companion, the 
proper motion of the central star should have an oscillating component on the order of $1.5-2$ au. 
If the orbit is significantly inclined, we should also observe a radial variance ($\sim 1.8\rm\, km/s$ if
the inclination is the same as the disk) with a period of about $20$ years. 

Note that the discussion on the secondary stellar companion and the displaced focus will
change the proper motion discussed earlier. The Keplerian velocity will change with a different 
total mass of the central binary system and with a different distance to the displaced focus. 
The system may not be super Keplerian anymore 
if the total stellar mass is larger. 
More detailed future analyses are needed to account for these additional complications. 

\subsection{Velocity maps of line emissions}
\begin{figure}[!htb]
\includegraphics[width=0.48\textwidth]{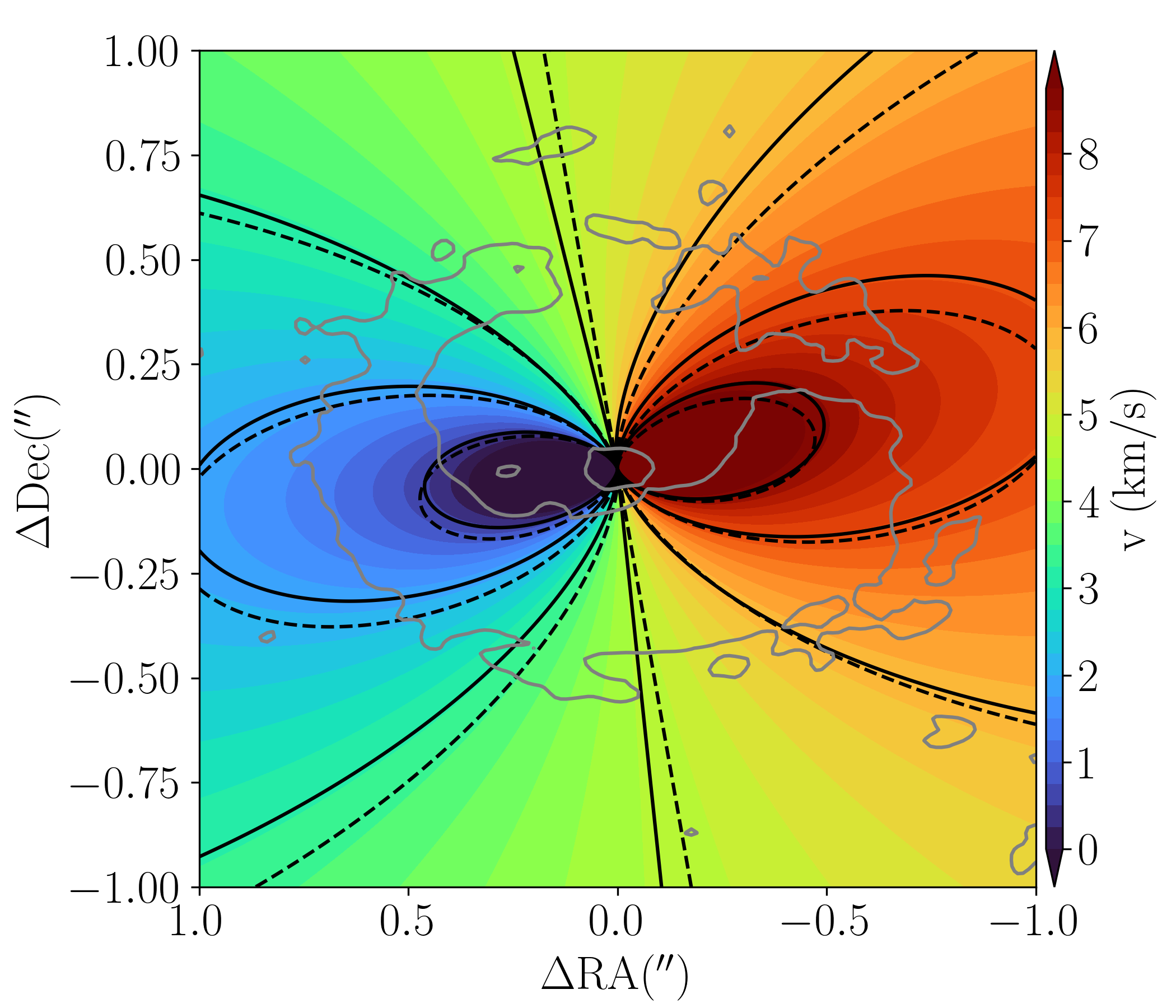}
\caption{The velocity map of the disk, assuming an elliptical orbit with $e=0.27$, as fitted from Sec.~\ref{ssec:ellipse}.
The colorbar is similar to \cite{vanderMarel2021_IRS48} with $v_\mathrm{source}=4.55\rm\, km/s$. 
The solid contours represent the same velocity map as the color map at the levels of 
$(-4, -2.67, -1.33, 0, 1.33, 2.67, 4)\rm\, km/s+v_\mathrm{source}$. 
The dashed contours are plotted at the same velocity levels but assumes circular (Keplerian)
orbit. A gray $3\sigma$ contour of the non-polarized flux from our observation is also
plotted as a background.}
\label{fig:vmap}
\end{figure}

The elliptical
orbit also has an impact on the velocity maps of line emission, assuming the gas is co-moving with the dust. 
The velocity map with an eccentricity of $e=0.27$ from our best-fit model is shown
in Fig.~\ref{fig:vmap}. 
We adopt a systemic velocity of 4.55 km/s as in \cite{vanderMarel2021_IRS48} and the resulting velocity map is similar to their Fig. 1. 
In order to compare with Keplerian
rotation, we also plot the iso-velocity contours for the elliptical orbit as black solid lines and the contours with same
levels for Keplerian rotation as black dashed lines. We can see that the red region is significantly
larger than the blue region, and 
the red-shifted solid contours are larger than their blue-shifted counterparts. 
This is in contrast with the circular orbit, where the dashed 
red-shifted contours are similar in sizes to the dashed blue-shifted contours. This is because the 
western part of the disk is closer to the perihelion 
and will show larger velocities in an elliptical orbit.
This asymmetry that the red-shifted contours are larger than their blue-shifted counterparts in velocity maps is visible in both $\rm H_2CO$ and $\rm ^{13}CO$ \citep{vanderMarel2014,vanderMarel2021_IRS48}, 
which lends support to the elliptical orbit (as first discussed by \citealt{Calcino2019}), although it remains to be determined whether the velocity map is quantitatively consistent with the eccentricity of $e=0.27$ inferred from the dust emission distribution.  

\section{Summary}
\label{sec:summary}

In this work, we present deep submillimeter ALMA observations towards the transition disk IRS 48 with a high spatial resolution. The
main findings are as follows:
\begin{enumerate}
\item For the first time, we are able to trace $95\%$ of the ring crossing the well-known crescent-shaped structure.
This ring is surprisingly eccentric with a large eccentricity of $0.27$. 

\item We detected compact emission at 15 $\sigma$ level that is centered in the star and spatially well separate from the crescent-shaped structure. 
The dust mass is estimated as about 
$1.5\times 10^{-8}\rm\, M_\odot$, or $0.005\rm\, M_\oplus$, if the emission is mm-sized dust thermal emission. 
We do not resolve the central emission with $0.1''$ beam. If the central object is an unresolved inner disk, the disk radius is a couple of astronomical units at most. 


\item 
We fit the proper motion of the crescent-shaped dust structure as $0.49\pm 0.26\, v_K$. 
Existing data do not support either circular orbit or elliptical orbit.
Detailed modeling and future high-resolution observations may help to understand 
the nature of the proper motion of the crescent peak.
\end{enumerate}

\section*{Acknowledgement}
We thank the referee for constructive reports that helped improve the manuscript significantly.
We thank Lile Wang, Greg Herczeg, Ruobing Dong, and Pinghui Huang for fruitful discussions.
HY is supported by the National Key R\&D Program of China (No. 
2019YFA0405100) and the China Postdoctoral Science Foundation (No. 2022M710230).
ZYL is supported in part by NASA 80NSSC18K1095 and NSF AST-1910106. 
LWL and REH acknowledge support from NSF AST-1910364. 
ZYDL acknowledges support from the Jefferson Scholars Foundation and the NRAO ALMA Student Observing Support (SOS) SOSPA8-003. 

This paper makes use of the following ALMA data: ADS/JAO.ALMA\#2019.1.01059.S, 2011.0.00635.SSB, 2013.1.00100.S, and 2017.1.00834.S. 
ALMA is a partnership of ESO (representing its member states), NSF (USA) and NINS (Japan), 
together with NRC (Canada), MOST and ASIAA (Taiwan), and KASI (Republic of Korea), in 
cooperation with the Republic of Chile. The Joint ALMA Observatory is operated by 
ESO, AUI/NRAO and NAOJ.

\bibliography{refs}{}

\end{CJK*}
\end{document}